\documentclass[aps,prl,twocolumn,superscriptaddress]{revtex4-1}

\usepackage{upgreek}		% Adds straight greek
\usepackage{graphicx}		
\usepackage{epstopdf}		
\usepackage{dcolumn}		
\usepackage{bm}				
\usepackage{bbold}			
\usepackage{float}			
\usepackage{natbib}			% For revtex
\usepackage{textcase}		% For revtex
\usepackage{titlesec}		% For revtex
\usepackage{url}			% For url bib

\usepackage[UKenglish]{babel}
\usepackage[utf8]{inputenc} 	%adds non-UK standard accents

\usepackage{color}  		 % May be necessary if you want to color links
\usepackage{hyperref}
\hypersetup{
    colorlinks=true, 			% set true if you want colored links
    linktoc=all,    		 	% set to all if you want both sections and subsections linked
    linkcolor=green,  		% choose some color if you want links to stand out
}

\graphicspath{{Images/}}

\makeatletter
\renewcommand{\fnum@figure}{ FIG. \thefigure}
\makeatother

\begin{document}

\author{R.~Saint}
\author{W.~Evans}
\affiliation{School of Physics and Astronomy, The University of Nottingham, Nottingham, NG7 2RD, United Kingdom}
\affiliation{Department of Physics and Astronomy, University of Sussex,
Brighton, BN1 9QH, United Kingdom}
\author{Y.~Zhou}	
\author{T.~M.~Fromhold}
\affiliation{School of Physics and Astronomy, The University of Nottingham, Nottingham, NG7 2RD, United Kingdom}
\author{E.~Saleh}
\author{I.~Maskery}
\author{C.~Tuck}
\author{R.~Wildman}
\affiliation{Faculty of Engineering, EPSRC Centre for Innovative Manufacturing in Additive Manufacturing, University of Nottingham, Nottingham, United Kingdom}
\author{F.~Oru\v{c}evi\'{c}}
\author{P.~Kr\"uger}
\affiliation{School of Physics and Astronomy, The University of Nottingham, Nottingham, NG7 2RD, United Kingdom}
\affiliation{Department of Physics and Astronomy, University of Sussex,
Brighton, BN1 9QH, United Kingdom}

\title{3D-printed components for quantum devices}
\date{\today}

\begin{abstract}

Recent advances in the preparation, control and measurement of atomic gases have led to new insights into the quantum world and unprecedented metrological sensitivities, e.g.\ in measuring gravitational forces and magnetic fields. The full potential of applying such capabilities to areas as diverse as biomedical imaging, non-invasive underground mapping, and GPS-free navigation can only be realised with the scalable production of efficient, robust and portable devices. Here we introduce additive manufacturing as a production technique of quantum device components with unrivalled design freedom, providing a step change in efficiency, compactness and facilitating systems integration. As a demonstrator we present a compact ultracold atom source using less than ten milliwatts power to produce large samples of cold rubidium gases in an ultrahigh vacuum environment. This disruptive technology opens the door to drastically improved integrated structures, which will further reduce power consumption, size and assembly complexity in scalable series manufacture of bespoke quantum devices.

\end{abstract}


\maketitle

\section{I. INTRODUCTION}
The current expansion in the field of quantum technologies and in particular quantum sensors has given rise to notable strides forward in device sensitivities with applications ranging from satellite independent navigation to non-invasive biomedical imaging \cite{Keil_FifteenYearsReview,Barrett_RotationReview}. As demand for wider use of these instruments grows, it is imperative to develop robust low-footprint components to provide practical portable systems. 

Recent work has focused on the exploitation of quantum phenomena in compact out-of-laboratory devices based on latest advances in micro-manufacturing technology including optical and electron beam lithography \cite{Nshii2013}, waveguide writing \cite{Hinds1998} and reactive ion beam etching \cite{Sewell2010}. 
These developments have enabled integrated atom chip based systems \cite{lewis2009fabrication,Folman2002} and commercial development of complete systems on a vehicle payload scale \cite{iSense2011,tino2013_spacecoldatoms,farkas2014_coldquanta}. Additive manufacturing (3D-printing) is a significant emerging technology capable of providing solutions to a range of problems due to the design freedom it permits \cite{AM_King,AM_Yilmaz} during the scalable production of individually bespoke components. In this work we demonstrate how this radically different manufacturing approach can be used to reach a step change in performance of practical scalable quantum devices.

 \begin{figure}[h]
\includegraphics[width=1\columnwidth]{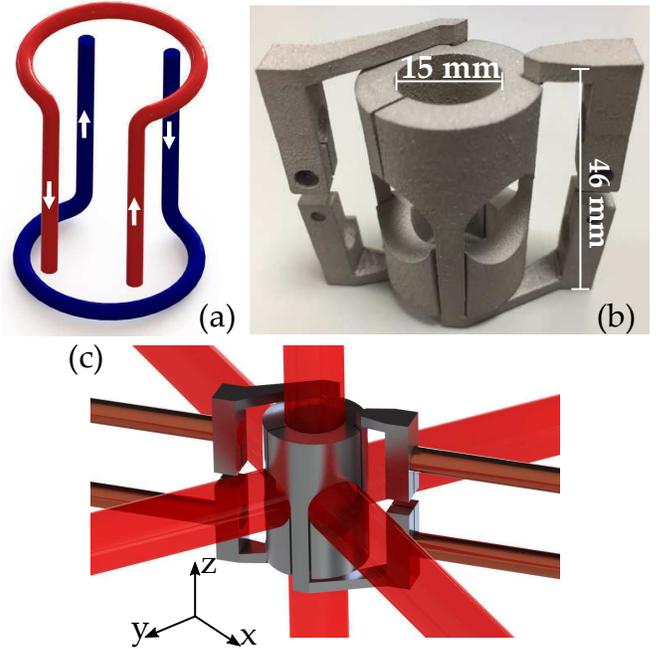}
\caption{(a) Schematic of the cylinder atom-trap current flow. (b) 3D-printed atom trap structure. (c) Digital render of the cylinder structure shown with vacuum feedthroughs and laser beams.}
\label{fig:schmetics_combi_image}
\end{figure}

Quantum resources are fundamental to all quantum technologies and here we present a centimetre scale demonstrator device providing such a resource through production of gases of ultracold atoms with unprecedented efficiency. Many attempts to reduce experimental demands in atom trapping have been considered including single-beam magneto-optical trap (MOT) designs based on conical \cite{Lee_PyramidConical}, pyramidal \cite{Xu_Pyramid} and tetrahedral \cite{Vangeleyn_Tetrahedral} diffraction-based reflectors to reduce optical equipment. Evolving atom chip technology is being pursued for small-scale field production \cite{Salim_CompactUltracoldChip,Keil_FifteenYearsReview,Rushton_Review}. While power consumption in the high-resistance chip conductors remains large, our approach in contrast allows us to trap tens of millions of atoms with merely $4\,\text{mW}$ power consumption. Our ultrahigh vacuum compatible device reliably produces a gas of up to $10^{8}$ rubidium-$87$ atoms at a temperature of $20\,\upmu\text{K}$. We report here the design, manufacture and characterisation of the key performance properties.
 
\section{II. DEVICE DESIGN} 
The operating principle of a MOT relies on a quadrupole magnetic field configuration.  Atoms are cooled, accumulated and trapped in the region of vanishing field where pairs of counter-propagating laser beams along three orthogonal spatial directions are overlapped at the same time. In the standard experimental implementation a pair of coils with parallel orientation and counter-propagating currents (anti-Helmholtz configuration) is used to form linear field gradients with a central field zero as required. While this type of set-up, normally assembled exterior to the vacuum chamber containing the cold atoms, is flexible and robust within a laboratory environment, power consumption remains high and portability is compromised.

It is possible to produce a planarised conductor configuration to form a quadrupole field with an out-of-plane zero \cite{Weinstein_NeutralTraps}. This allows for standard chip implementation. Such two-dimensional (2D) designs, however, are at a disadvantage in their power-efficiency when compared to three-dimensional implementations. This is because the out-of-plane zero must be formed by multiple in-plane currents, whose fields compensate each other at the zero position. As all these fields necessarily drop monotonically in magnitude with distance from the plane, the field {\em gradients} will also at least partially compensate each other at the field zero. This gradient compensation needs to be minimised in order to obtain a power-efficient planar solution. In contrast in three dimensions (3D) the currents generating the fields that cancel at the trap centre position of a device can be formed in such a way that the gradients produced by them add, rather than subtract as in planar implementation. This argument can be generalised to all use of magnetic fields in trapping and manipulation of cold atoms, where field minima need to be created away from field-producing structures.

\begin{figure}[h]
\includegraphics[width=1\columnwidth]{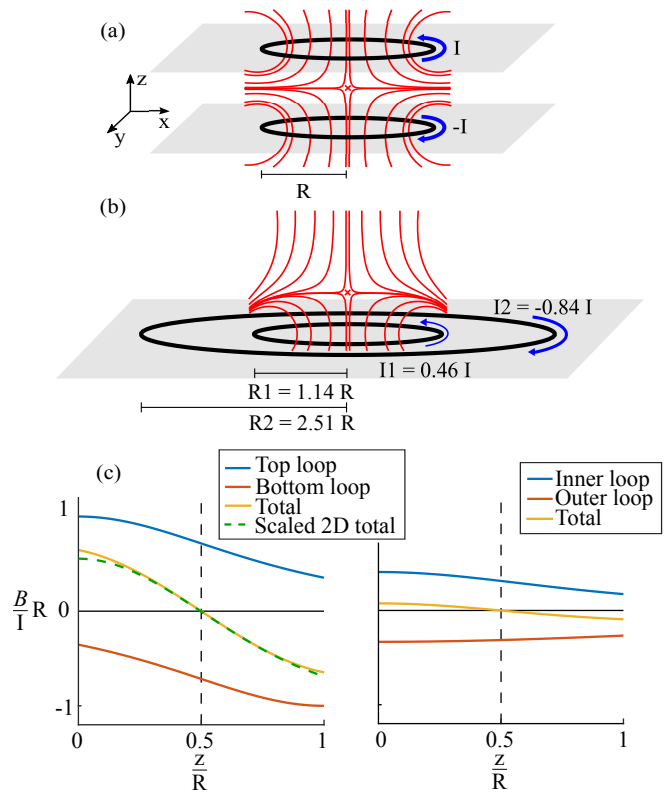}
\caption{Comparison of two quadrupole field generating structures whereby two current loops are placed in two parallel planes (3D) or a single plane (2D). Streamline plots show the field produced from the 3D system (a) and 2D system (b). (c) Magnetic fields $B_z$ along the symmetry axis $z$ of the current loops (due to symmetry radial fields vanish). The field of one loop (blue) is compensated by the other (red) in both configurations at the same zero-field position [red cross (x) in (a) and (b)]. At equal power consumption, the total field gradient (yellow) for the 3D system is stronger than that in the 2D case by a factor of $7.37$. To reach the same gradient with the planar 2D assembly, the current needs to be scaled up by the same factor (dashed green line). This corresponds to an increased power dissipation by a factor of 54.3.}
\label{fig:gradient_calcs}
\end{figure} 

A simple idealised comparison of a 3D versus 2D scenario of generating a quadrupole field reinforces the above point. As a model, we choose two infinitesimally thin current loops in the anti-Helmholtz configuration (two parallel loops with equal radius $R$ carrying equal currents $I$ are placed at a distance $d=R$ from each other) and compare them to two in-plane concentric current loops (Figure~\ref{fig:gradient_calcs}). The in-plane loops have radii $R_1$ and $R_2$ and carry currents $I_1$ and $I_2$, respectively. The 3D configuration field zero occurs at a distance $R/2$ from either loop. Consequently, the planar configuration parameters are chosen such that a field zero forms at $R/2$ from the current plane in that case. We further impose equal power consumption in both configurations which is obtained when $R_1 I_1^2 + R_2 I_2^2= 2 R I^2$. Finally, we require the field curvature to vanish at the field zero, so that in approximation of the ideal quadrupole field the field varies only linearly at this centre position. Note that without loss of generality it is sufficient to only consider the field along the loops' symmetry axis $z$, where the field is always oriented along $z$. We find that under the above constraints the maximal gradient is achieved in the planar configuration when $R_1 = 1.14\,R$ and $R_2 = 2.51 \, R$ with the currents $I_1 = 0.46 \, I$ and $I_2 = -0.84 \, I$. Even in this optimal configuration, the gradient is reduced by a factor larger than $7$ with respect to the 3D anti-Helmholtz configuration. Calculated field configurations are shown in Figures \ref{fig:gradient_calcs}(a) for a 3D structure and (b) for a planar structure. Figure \ref{fig:gradient_calcs}(c) displays the corresponding fields along the symmetry axis of the loops. To match the gradient obtained in the 3D configuration with the 2D configuration requires a more than 50-fold increase of power consumption.

Miniaturisation of components generally has the advantage of reducing power consumption. Again, for the 3D case, the idealised anti-Helmholtz configuration may serve as an illustration of a scaling law that is extendible to more general magnetic field generating structures. In the anti-Helmholtz configuration, as in other configurations e.g.\,for magnetic traps of various shapes, the key parameter is the generated field {\em gradient} at the field zero (or generally: field minimum). As the field of a single current loop at the position of the quadrupole field zero ($z=R/2$) scales as $\sim 1/R$, the gradient at that position scales as $1/R^2$. Conversely, in order to maintain a constant gradient, the required current scales as $\sim R^2$. If the conductor cross section is assumed to scale with size of the device, the resistance $Z$ of the structure increases with the length of the current loop ($\sim R$) and drops as $\sim 1/R^2$ with the cross section, such that the Ohmic power dissipation overall scales as $P=ZI^2\sim R^3$, i.e.\ with the {\em volume} of the device. Typically, a design goal is to reduce power, especially for portable device implementations. It should be noted that for heat management, sometimes {\em power density} is primarily relevant, which is size independent. At a given device size, optimal use of the available volume for conductors reduces both power and power density.

\section{III. PROTOTYPE}
All the above considerations show that an ideal device should be based on miniaturised 3D structures with optimal use of the available volume, which is typically defined by other constraints, such as access to laser beams for atom cooling and detection in our demonstrator device. Thus far, these design rules could not be implemented as common devices relied on either larger-scale 3D structures without sufficient design freedom for optimisation \cite{Wil2004} or on miniaturised complex 2D chip-geometries \cite{Keil_FifteenYearsReview}. Additive manufacturing for the first time allows for a robust and scalable combination of these two approaches and hence makes full and simultaneous implementation of the above design rules possible, as the specific model case of our demonstrator illustrates.

A miniaturised actual low-power design of an electric current path resulting in a quadrupole field pattern can be provided as shown schematically in Figure \ref{fig:schmetics_combi_image}(a). A set of central parallel straight conductors carrying alternating currents forms a two-dimensional quadrupole field. The translational symmetry is broken by counter-propagating current loops in two parallel planes orthogonal to the four straight conductors. 

We use computer aided design (CAD) to create an experimentally viable design with a sizeable trapping region, optical access, structural supporting arms and electrically conductive contacts. Fitting these constraints we produce a centimetre scale cylindrical device [Figure \ref{fig:schmetics_combi_image}(b)] forming a trapping region of $\sim(5\,\text{mm})^3$. The overall volume of the device is chosen to be small for reasons of power consumption while still being large enough to permit laser beams with sufficient diameter to enter the central trapping region. We include electric feeds that simultaneously serve as supporting arms with clamping connections to commercially standard CF40 electrical feedthroughs. We allow for insulation between the two opposing currents by manufacture of two separate monolithic components. The design volume is maintained where possible to maximise current carrying material and mitigate resistive heating. The final printed structure is displayed in Figure~\ref{fig:schmetics_combi_image}(b) and a digital render of the experimental system is shown in Figure~\ref{fig:schmetics_combi_image}(c).

\begin{figure}[h]
\includegraphics[width=1\columnwidth]{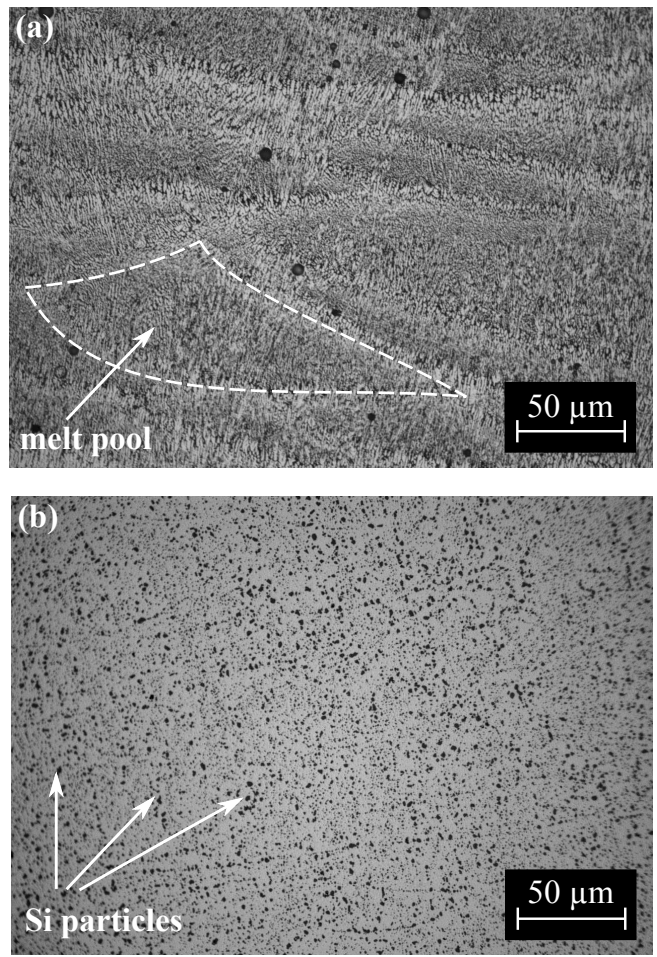}
\caption{Typical surface pattern images from an optical microscope. (a) As-built structure showing the material immediately after manufacture; the darker regions separated by lighter channels are melt pools formed by the laser during SLM. (b) Images following solution heat treatment showing the more uniform distribution of silicon.}
\label{fig:heat_treatment}
\end{figure}

The additive manufacturing technique we employ for 3D-printing this device is selective laser melting of an Al-Si10-Mg alloy \cite{Maskery_PrintingDetails}. This method is based on scanning a high-power laser over a bed of powdered metal. In addition to the standard technique, we apply solution heat treatment (SHT), which is usually carried out to improve mechanical properties. Here it enhances the electrical conductivity through significant changes in the microstructure, as shown in Figure \ref{fig:heat_treatment}. The resulting conductivity reaches 70\% of the bulk value, making heating effects in the material negligible in all our experiments. Further details are found in the methods section.

The device manufactured in this way is compatible with ultra-high vacuum conditions. An off-the-shelf compact vacuum system houses the cylinder trap and is pumped down to $\sim 10^{-7}\,\text{mbar}$ with a turbomolecular pump. Ultra-high vacuum ($ <10^{-10}\,\text{mbar})$ is then achieved after baking the chamber at $200\,^{\circ}\text{C}$ for a week and pumping via a passive non-evaporable getter pump and a compact $6\,\text{l/s}$ ion pump \cite{NexToRR}. UHV conditions were maintained for the nine month duration of experiment with no detectable outgassing, an unprecedented result in the vacuum compatabiltiy of 3D-printed Al-Si10-Mg. Titanium and silver have shown similar capabilities \cite{Ashley_VacuumCompatibilitySilerTitanium,Povilus_3DPrinted_VacuumStuff}. 

\section{IV. RESULTS}
\begin{figure}[h]
\includegraphics[width=1\columnwidth]{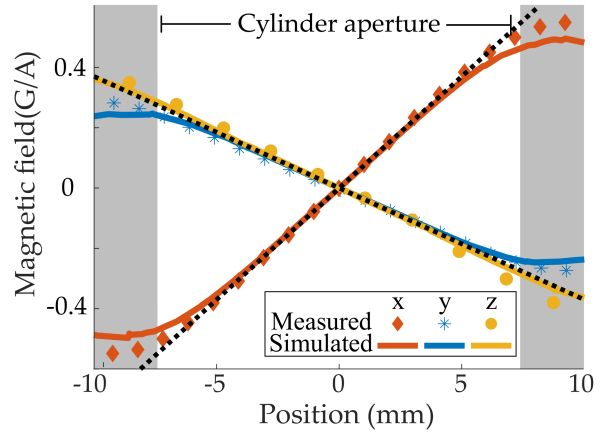}
\caption[width=1\columnwidth]{Plot of the simulated (finite-element) versus measured magnetic field magnitude along each axial direction. Dotted lines here show ideal -2:1:1 ratio between strong and weak axes.}
\label{fig:sim_measured_plots}
\end{figure}

A trapping field gradient of $\sim10\,\text{G/cm}$ along the strong axis is typical when using rubidium-87. Field gradients across the trapping region are linear in the ideal case. Finite element simulations of field production were conducted to aid design and the simulated magnetic field profiles are shown in Figure \ref{fig:sim_measured_plots} alongside measured values of the field strength. Clearly the fields produced by our prototype closely fit both criteria and the field geometry produced is that expected of the anti-Helmholtz configuration with a -2:1:1 ratio between strong and weak axes. A gradient of $10\,\text{G/cm}$ is produced at a manageable current of $15\,\text{A}$. Measurements at currents up to $50\,\text{A}$ with corresponding gradients of up to $40\,\text{G/cm}$ have been performed, confirming the expected linear relation between gradient and current and the wide range of control of the device within the regime of negligible heating.

We can quantify the performance of our cold-atom source by correlating the power consumption of the device to the achieved atom number. The required cloud size is dependent on system use. In our demonstration experiments we aimed to produce $10^{7}$ atoms corresponding to the standard set in existing thermal atom quantum devices \cite{iSense2011}. We find that our  device is even capable of producing a cloud of more than $10^{8}$ atoms as necessary for Bose-Einstein condenstate experiments \cite{Rushton_Review}. Measurements are taken of the atom number, via absorption imaging, for varying cylinder currents $I$ and corresponding power consumptions $P=Z I^2$ with the Ohmic device resistance $Z$.

First we consider the capabilities of the device with a maximal beam diameter of $15\,\text{mm}$. Atom numbers exceeding $10^8$ are reached for power levels of $30\,\text{mW}$ and above (the typical gradient of $10\,\text{G/cm}$ corresponds to $\sim 150\,\text{mW}$), with a broad plateau with atom numbers dropping by $50\%$ at $10\,\text{mW}$ power consumption and below. Even down to the lowest measured power of $4\,\text{mW}$ the atom number stayed well above $10^7$. With regards to standard atom traps, such as coil systems mounted externally to the chamber or macroscopic intra-vacuum conductor assemblies in BEC experiments, which can require up to $10\,\text{W}$ of power, a power reduction of orders of magnitude is found for our prototype. This confirms the scaling of power with device volume discussed above, i.e.\,the cubic scaling of power with a characteristic length scale of the device.

The above conceptual considerations supported by the experimental findings obtained with a small volume device suggest that further miniaturisation could lead to ever decreasing power consumption. It has, however, been found that laser cooling with small beam diameters becomes inefficient. The exact scaling of the atom number N with beam diameter is largly dependent on the experimental system, but strong power laws have been verified, e.g.\ $N\sim D^{5.82\pm0.05}$ \cite{Camara2014}. Leaving clear apertures for sufficiently large laser beams is therefore a constraint that limits the drive towards smaller volume trapping assemblies. The size of the optimal apertures depends on the target atom number used in e.g. the employed quantum sensing scheme. For our prototype we chose the $15\,\text{mm}$ diameter aperture to guarantee atom numbers in excess of $10^7$ while still keeping power levels in the lower mW range. 

\begin{figure}[h]
\includegraphics[width=1\columnwidth]{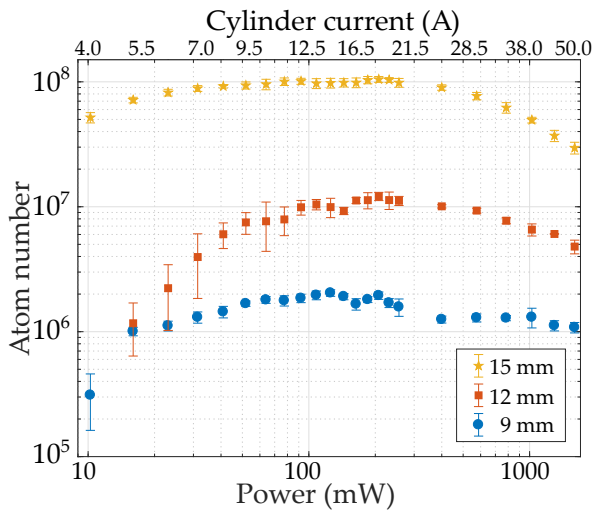}
\caption{Maximum atom number as a function of cylinder power consumption (current) for three beam diameters. The current range $4\,\text{A to }50\,\text{A}$ corresponds to the magnetic field gradient range $3.2\,\text{G/cm to }40\,\text{G/cm}$.}
\label{fig:AtomNum_Power}
\end{figure}

In order to empirically explore whether even lower apertures are viable and consequently further strong power reduction (scaling with the cube of the aperture), we performed measurements with reduced cooling laser beam diameters (Figure~\ref{fig:AtomNum_Power}). Following a $20\%$ reduction of the beam diameter from $15\,\text{mm}$ to $12\,\text {mm}$ the device still yields of more than $10^{7}$. In an aperture size adjusted prototype this would now be achieved at $50\%$ of the original power level (in our simulated case we use $\sim 200\,\text{mW}$). A further beam diameter reduction to $9\,\text{mm}$ ($60\%$ of the original size) allows for atom numbers of $(2.0\pm0.1)\times10^{6}$. A size-adjusted device would produce the same gradients as our manufactured prototype at merely $20\%$ of the power, in this case. Full measurements for the currents and power levels actually used in our system (ranging between $10\,\text{mW}$ and $1.6\,\text{W}$) are shown in Figure \ref{fig:AtomNum_Power}. For each cylinder current the cooling light is detuned to maximise atom number in accordance with the dependence of the applied current to the field gradient \cite{Wineland_LaserCooling}. The peak red-detuning varies from $16\rightarrow25\,\text{MHz}$ for currents $4\rightarrow25\,\text{A}$. In all cases the atom number reaches a plateau at currents of $\sim10\,\text{A}$, corresponding to field gradients along the strong quadrupole axis of $\sim 7\,\text{G/cm}$. Operation at high atom numbers is possible over a wider range of gradients ($\sim7-20\,\text{G/cm}$) with only gradual reduction for smaller (and larger) gradients.

\begin{figure}[h]
\includegraphics[width=1\columnwidth]{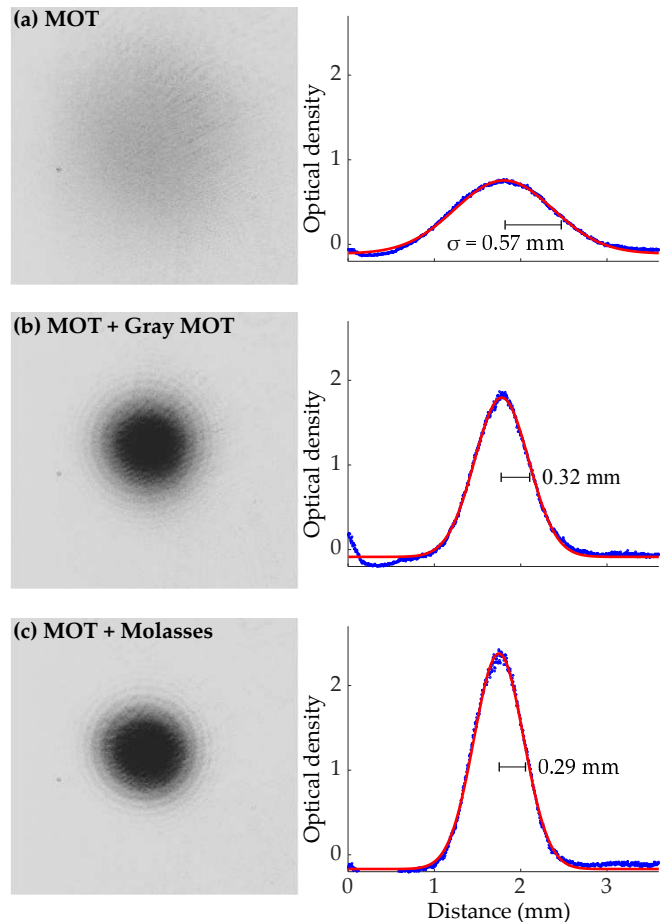}
\caption{Left: Optical density images of cloud of atoms in three cooling regimes, each taken after $12\,\text{ms}$ time of flight. Right: Vertically integrated optical densities (blue) with superimposed Gaussian fits (red). (a) The typical MOT cloud, (b) MOT cloud after gray MOT cooling (see text), (c) MOT cloud after molasses cooling.}
\label{fig:Grey_MOT_Schematic}
\end{figure}

Finally, a key property of a useful quantum resource is the temperature of the cloud as colder samples typically enhance the performance of sensors. With usual trapping techniques we produce a MOT with a temperature of  $170\pm 4\,\upmu\text{K}$ as a result of the laser cooling during trapping; this temperature is close to the Doppler cooling limit for rubidium, $143\,\upmu\text{K}$. Temperatures are measured using time-of-flight expansion \cite[4.5]{Ketterle_BECs}. An example optical density image of our MOT can be seen in Figure \ref{fig:Grey_MOT_Schematic}(a). Figure \ref{fig:Grey_MOT_Schematic}(b) is an example optical density image taken following a simpler `Gray MOT' cooling sequence \cite{Fernandes2012} carried out at the end of MOT loading. This scheme, in which the magnetic fields are left active during the cooling, is robust with respect to optical alignment and ambient magnetic field control. A ramp of $6\,\upmu\text{s}$ to a maximal red-detuning of $10\Gamma$ ($\Gamma$ being the natural line width of the transition = $6.065\,\text{MHz}$) is used here to yield a cloud of $(3.7\pm0.2)\times10^{7}$ at a temperature of $29.1\pm1.2\,\upmu\text{K}$. Even lower temperatures can be achieved by sub-Doppler cooling implemented with a molasses cooling scheme involving a $3.2\,\upmu\text{s}$ ramp to a maximal red-detuning of $10\Gamma$ after MOT loading, with no magnetic fields. Figure \ref{fig:Grey_MOT_Schematic}(c) shows a cloud of $(4.0\pm0.2)\times10^{7}$ atoms at a temperature of $(20.1\pm0.2)\,\upmu\text{K}$ as a result of applying this scheme. 

\section{V. CONCLUSION} 
We have established a viable technique for the design and manufacture of vacuum compatible 3D-printed atom trapping modules that are ready to function as the core quantum resource within a wide range of quantum devices. Exploiting the power reduction granted by producing fields close to the trapping region and the preferential  three-dimensional current density we have produced an extremely low power device operating with as little as $4\,\text{mW}$. The ability to produce $>10^{7}$ cold atoms makes this technique a prime candidate to replace the standard quantum resource in thermal atom systems. The further  capability of providing large samples of $>10^{8}$ cold atoms paves the path towards a simple quantum gas source in devices using Bose-Einstein condensates (BECs) as resource. The extreme reduction in power consumption of the trapping structure minimises the required resources for the system as a whole and with the rapidly progressing technological developments of associated experimental systems and evolving 3D-printing techniques significant further power reduction and system integration is expected. Future designs could incorporate mounting structures for optical components following trends in minimised optics systems or allow for integration with 2D-MOT systems or atom chip technology to allow for alignment with BEC experimental procedures. 

\section{ACKNOWLEDGEMENTS}
We acknowledge support of the Engineering and Physical Sciences Research Council (EPSRC), the UK National Quantum Technology Hub for Sensors and Metrology and the Defence Science and Technology Laboratory (Dstl) as well as funding through a University of Nottingham Discipline Bridging Award.

\section{APPENDIX A: PRODUCTION DETAILS} A Renishaw AM250 selective laser melting (SLM) machine is used to produce Al-Si10-Mg samples from a powder-alloy of chemical composition Al $88.9\,\text{wt}\%$, Si $10.7\,\text{wt}\%$, Mg $0.5\,\text{wt}\%$ (particulate sizes $15\,\upmu\text{m}$ to $100\,\upmu\text{m}$) \cite{Maskery_PrintingDetails}. Structures are created by melting the successive layers of powder with a 200 W Yb-Fiber ($\lambda = 1064\,\text{nm}$) laser. Current 3D-printing methods are capable of manufacture with various alloys of titanium, steel, stainless steel and silver. Here Al-Si10-Mg is used for the convenient electrical properties and low cost. 

Once printed, the structure is heated at $520\,^{\circ}\text{C}$ for 1 hour, followed by water quenching and artificial aging at $160\,^{\circ}\text{C}$ for 6 hours with both steps carried out in a pre-heated furnance with an air atmosphere \cite{Aboulkhair_HeatTreatment}.This process yields a cold resistivity of $5\times 10^{-8}\,\Omega \text{m}$  or a resistance of $640\pm 4\,\upmu\Omega$ for our geometry which is a $20\%$ reduction from the as-printed structure resistance of $800\pm 20\,\upmu\Omega$. A direct comparison of pure aluminium \cite{KayeLaby} puts the  conductivity of this Al-Si10-Mg alloy at $70\%$ of the value for bulk material at room temperature. The dominant cooling mechanism of the mounted device is thermal conduction through the electric feedthrough simultaneously serving as mechanical mount. The body of the vacuum chamber serves as heat sink. As convection (air cooling) is irrelevant in this set up, it is sufficient to measure heating effects under atmospheric conditions. We performed measurements for currents of up to $50\,\text{A}$, with the hottest-point temperature (located at the corner of the cylinder arm) never exceeding  $36\,^{\circ}\text{C}$.

\section{APPENDIX B: LIGHT FREQUENCY MANIPULATION} Our optical cooling field is generated by six independent beams created by a diode laser locked to the $^{87}$Rb D2 line, $5^{2}S_{1/2} \rightarrow 5^{2}P_{3/2}$ hyperfine \textit{cooling transition} ($\text{F}=2 \rightarrow \text{F'}=3$) providing a per-beam power of $40\,\text{mW}$ distributed across Gaussian beam with $1/\text{e}^{2}$ diameters of $2\,\text{inches}$ ($50.8\,\text{mm}$), which are truncated with iris diaphragms to the diameter of the optical access ports of the atom trap. The light frequency is red-detuned by $20\,\text{MHz}$ ($\sim$ three times the natural line width) from the cooling transition. 

To close the transition, a small amount of additional light at the \textit{repumping transition} $\text{F}=1 \rightarrow \text{F'}=1$ is required. We derive this light from a separate diode laser, whose light is coupled into one of the three fibre pairs feeding the light to the trapping region. A resonant imaging beam $(\text{F}=2 \rightarrow \text{F'}=3)$ is then coupled into one of the vertical beams ($z$-axis) to generate absorption images \cite[3.2]{Ketterle_BECs} of the various atom clouds. Our fully fiberised set up is immediately ready for a replacement of our commercial laser sources by integrated miniature systems as are becoming readily available at the moment \cite{QTHub}. 

\section{APPENDIX C: LIAD RUBIDIUM DESORPTION}
A $^{87}$Rb background pressure is generated via current flow through evaporative dispensers. Aditionally, a light induced atomic desorption (LIAD) \cite{Klempt_LIAD} ultra-violet LED system is triggered in the experimental sequence. By using LIAD only during the loading phase of our MOT we are able to achieve a lower average background pressure of rubidium in the chamber, as well as reducing the required dispenser current. This process can also improve the lifetime of any magnetic traps added to later designs. 

\section{APPENDIX D: EXPERIMENTAL CONTROL SYSTEM}
The control systems of cold atom experiments have traditionally relied on expensive hardware running bespoke user created software or scripts developed in environments such as C++ and LabVIEW which require significant expertise in software development. To provide a low-power, small-volume ($1.2\text{l}$) and low-cost portable solution, our experimental control is implemented in a master-slave configuration comprising of three micro-controllers.  One \textit{master} micro-controller provides 12 fast digital transistor-transistor logic (TTL) signals, two of which are used to communicate to two \textit{slave} devices, each providing two analogue channels. Simple pin manipulation techniques are used in place of standard development environment functions to reduce the minimum on-off pulse width from $5\,\upmu \text{s}$ to $100\,\text{ns}$. A compact battery array can service the three $5\,\text{V}$ micro-controllers and attached simple $12\,\text{V}$ step-down and amplification circuits. The low-cost hardware and accessible high-level integrated development environment (IDE) allows for fast and flexible design and prototyping of experimental sequences for the trapping and cooling of atoms. 

\bibliographystyle{apsrev4-1}
\bibliography{Bibliography}

\end{document}